\documentclass[onecolumn]{article}
\usepackage{graphicx}
\usepackage{amsfonts}
\usepackage{amssymb}
\usepackage{amsmath}
\usepackage{dcolumn}

\input{tcilatex}
\begin{document}

\title{APPROXIMATION FOR A TOY \emph{DEFECTIVE} ISING MODEL}
\author{Adom Giffin\thanks{%
E-mail: physics101@gmail.com} \\
%EndAName
Princeton Institute for the Science and Technology of Materials\\
Princeton University\\
Princeton, NJ 08540, USA}
\maketitle

\begin{abstract}
It has been previously shown that one can use the ME methodology (Caticha
Giffin 2006) to reproduce a mean field solution for a simple fluid (Tseng
2004). One could easily use the case of a simple ferromagnetic material as
well. The drawback to the mean field approach is that one must \emph{assume}
that all atoms must all act the same. The problem becomes more tractable
when the agents are only allowed to interact with their nearest neighbors
and can be in only two possible states. The easiest case being an Ising
model. The purpose of this paper is to illustrate the use of the ME method
as an approximation tool. The paper show a simple case to compare with the
traditional mean field approach. Then we show two examples that lie outside
of traditional methodologies. These cases explore a ferromagnetic material
with \emph{defects}. The main result is that regardless of the case, the ME
method provides good approximations for each case which would not otherwise
be possible or at least well justified.
\end{abstract}

\section{Introduction}

In a simple ferromagntic material (single domain), the electronic spin of
the individual atoms are strong enough to affect one and other, the so
called "exchange" interaction \cite{Reif}. However this effect is temperture
dependent. When the temperture is below a certain point (the Curie
tempreture) the spins tend to all point in the same direction due to their
influence on each other. This establishes a permanant magnet as the
individuals produce a net dipole effect. Above this temperture, the atoms
cease to have a signifigant effect on each other and the material behaves
more like a paramagnetic substance. Determining this net dipole effect can
be difficult. First, the interactions are due to complicated quantum
effects. Second, since a given material as a very large number of atoms,
computing the net dipole effect can be difficult in two dimensions and
completely intractable in three dimensions. Therefore approximations are
made to facilitate computational difficulties such as using an Ising Model
and or the mean field approximation.

In 1957, Jaynes \cite{Jaynes} showed that maximizing statistical mechanic
entropy for the purpose of revealing how gas molecules were distributed was
simply the maximizing of Shannon's information entropy \cite{Shannon1948}
with statistical mechanical information. This idea lead to MaxEnt or his use
of the Method of Maximum Entropy for assigning probabilities. This method
has evolved to a more general method, the method of Maximum (relative)
Entropy (ME) \cite{CatichaGiffin} which has the advantage of not only
assigning probabilities but \emph{updating} them when new information is
given in the form of constraints on the family of allowed posteriors. One of
the drawbacks of the MaxEnt method was the inability to include data. When
data was present, one used Bayesian methods. The methods were combined in
such a way that MaxEnt was used for assigning a prior for Bayesian methods,
as Bayesian methods could not deal with information in the form of
constraints, such as expected values.

Previously it has been shown that one can use ME to reproduce a mean field
solution for a simple fluid \cite{Tseng}. The purpose of this was to
illustrate that in addition to updating probabilities, ME can also be used
an an approximation tool. The purpose of our paper is to illustrate the use
of the ME method as a tool for attaining approximations for ferromagnetic
materials that lie outside the ability of traditional methods. In doing so
we further the the previous work done and show the versitility of the method.

In the second section of this paper we show the traditional methods for
attaining an approximation, the Ising model and the mean field field
approximation. The third section shows how ME can not only reproduce the
mean field approximation, but does so using fewer assumtpions. In the fourth
section we provide two examples that lie outside the capabilities of
traditional methodologies. These cases explore a ferromagnetic material with 
\emph{defects}. The main result is that regardless of the case, the ME
method provides good approximations for each case which would not otherwise
be possible or at least well justified.

\section{The Ising model and the mean field approximation}

The drawback to the mean field approach is that one must \emph{assume} that
all atoms must all act the same. Additionally, the problem is only tractable
when the atoms are only allowed to interact with their nearest neighbors and
assumed to be in only two possible states. This model is called the Ising
model.

\subsection{The Ising Model}

Althought these exchange interactions are quantum related, Ernst Ising \cite%
{Schroeder} suggested that one should examine a simple model where the
atomic spins are related by simple spins of +1 and -1 or spin "up" and spin
"down". Further, he suggested to neglect all exchange interactions \emph{%
except} those between nearest neighbors. Using these two assumptions, the
exchange interaction energy can be calculated as follows:%
\begin{equation}
H=-J\sum_{i,j}^{N,n}s_{i}s_{j}~,  \label{Hamiltonian_N}
\end{equation}%
where $N$ is the number of atoms, $J$ is the exchange interaction energy and 
$s$ is the net electronic spin of the atom where $s=\{+1,-1\}$. The standard
partition function for this system would then be written as, 
\begin{equation}
Z=\sum_{\left\{ s_{i}\right\} }e^{-\beta H}~,  \label{PartitionF_N}
\end{equation}%
where $\sum_{\left\{ s_{i}\right\} }=\sum_{s_{1}}\sum_{s_{2}}\ldots
\sum_{s_{N}}$ and $\beta $ is usual thermodynamic value $1/kT$ where $T$ is
the temperture and $k$ is the Boltzmann constant. For the one dimentional
case, the partition function can easily be calculated, 
\begin{equation}
Z=2^{N}(\cosh \beta J)^{N-1}\approx (2\cosh \beta J)^{N}\text{ (for large }N%
\text{).}  \label{Partion_I}
\end{equation}

The resulting average energy would then be, 
\begin{equation}
\bar{H}=-\frac{\partial \ln Z}{\partial \beta }=-NJ\tanh \beta J~.
\label{AvgI}
\end{equation}

\subsection{The mean field approximation}

Although the Ising model's partition function can be solved exactly, this is
only true for one dimension. It fails for real ferromagnetic substances
since it only takes into account two near neighbors. For three dimensional
materials, each atom may have 6, 8 or even 12 close neighbors depending on
the crystal geometry. This is where another approximation is made and it is
commonly refered to as the mean field approximation \cite{Reif, Schroeder}.

We start with the same energy function as above (\ref{Hamiltonian_N}) but
now examine only one atom (sometimes called the "central" atom), 
\begin{equation}
H_{i}=-Js_{i}\sum_{,j}^{n}s_{j}~.
\end{equation}%
Where $H_{i}$ is the energy for one atom and $n$ is the number of nearest
neighbors (ex. a face-centerd cubic lattice would have 12) with atoms at the
edge of the material neglected. Now we assume that each neighbor contributes
equally and \emph{average} over the neighbor spins around this one atom, 
\begin{equation}
\bar{H}_{i}=-Js_{i}n\bar{s}_{j}~,
\end{equation}%
where $\bar{s}_{i}=\sum_{,j}^{n}\frac{s_{j}}{n}$ and the Boltzmann
probablity for finding this atom in the $s_{i}$ state is 
\begin{equation}
p_{i}=\frac{1}{Z_{i}}e^{-\beta H_{i}}~,
\end{equation}%
where $Z_{i}=2\cosh \beta Jn\bar{s}_{j}.$ Now we take the the average of for
this one atom over its two possible states, 
\begin{equation}
\left\langle s_{i}\right\rangle =\sum_{s_{i}}s_{i}~p_{i}=\frac{1}{Z_{i}}%
\left[ (1)e^{\beta H_{i}}+(-1)e^{-\beta H_{i}}\right] =\frac{2\sinh \beta Jn%
\bar{s}_{j}}{2\cosh \beta Jn\bar{s}_{j}}=\tanh \beta Jn\bar{s}_{j}~.
\label{Expected_M}
\end{equation}%
The next assumption is that all of the \emph{all} atoms will behave like
this one. Therefore, the average spin for the entire material is 
\begin{equation}
\bar{s}=\left\langle \left\langle s_{i}\right\rangle \right\rangle
=\left\langle \bar{s}_{j}\right\rangle ~,
\end{equation}%
which now allows us to write 
\begin{equation}
\bar{s}=\tanh \beta Jn\bar{s}~,  \label{AvgM}
\end{equation}%
which is the final result for this section. This equation can then be used
to solve for the Curie temperture, the total magnetic moment and the
magnetic susceptibility. To be more general, an external magentic field can
be applied as well which would alter the solution but not the form. For this
case, the Hamiltonian would be, 
\begin{equation}
\bar{H}_{i\text{eff}}=-Js_{i}n\bar{s}_{j}~-\mu Bs_{i}=-\epsilon s_{i}~,
\label{Hamiltonian_effM}
\end{equation}%
where $\mu $ is the magnetic moment, $B$ is the magnetic field strength and $%
\epsilon $ is the net effective energy on the $i^{th}$ atom. Note however
that while this does not alter the solution using the mean field
approximation becasuse this is the effective Hamiltonian (\ref%
{Hamiltonian_effM}) for a single atom. In other words, the averaging is
already taken into account in the effective Hamiltonian. If one were to
write the full Hamiltonian with the external field instead of (\ref%
{Hamiltonian_N}) above, then one will not have as simple of a solution for (%
\ref{Partion_I}), i.e. $Z_{\text{eff}}\neq (2\cosh \beta \epsilon )^{N}.$

\section{Using ME for the approximation}

Part of the strength of using ME is that one does not need to justify some
of the \emph{assumptions} such as those used above (4 explicitly used, many
hidden). One simply supplies the information that one has available and
allows the method to turn out the most honest solution based on the
information given. This can be demonstrated by using the method to find an
appropriate approximation for a ferromagnetic material given the information
that one has available. We show in this section where ME was used for this
very same task to arrive at a similar solution as above.

We start by writing the full posterior solution for a ferromagnetic \emph{%
system} (all atoms) which is a canonical distribution, 
\begin{equation}
p\left\{ s\right\} =\frac{1}{Z}e^{-\beta H\left\{ s\right\} }~,  \label{P}
\end{equation}%
where $p\left\{ s\right\} =p(s_{1}\ldots s_{N}),$ $Z$ is the patition
function from (\ref{PartitionF_N}) with $H\left\{ s\right\} $ being some
general Hamiltonian and $N$ the number of atoms. NOTE: This function can be
assertained by using the ME method without assumptions as Jaynes did for
Gibbs' canonical function when he used his special case of ME, MaxEnt \cite%
{Jaynes}. However, to determine the values of the Curie temperture, the
total magnetic moment and the magnetic susceptibility we are faced with the
same dilema as above; we cannot compute the solutions exactly. Therefore, we
must try to find an apporximation that is tractable. For \emph{illustrative}
purposes, we will adopt the Ising model so as to show how one can use ME to
arrive at a similar approximation as the mean field produces. For this we
let 
\begin{equation}
H\left\{ s\right\}
=H_{int}+H_{ext}=-J\sum_{i,j}^{N,n}s_{i}s_{j}-\sum_{i}^{N}\mu Bs_{i}~,
\label{Hamiltonian_NF}
\end{equation}%
where in $H_{int}=H$ from (\ref{Hamiltonian_N}) and $H_{ext}$ is the energy
due to external magnetic field acting on the atoms. We now wish to use ME to
find an approximation that is tractable. We accomplish this by first writing
down the appropriate entropy, 
\begin{equation}
S[p_{\text{A}}|p]=-\sum_{\left\{ s\right\} }p_{\text{A}}\log \frac{p_{\text{A%
}}}{p}~,  \label{Entropy_ME}
\end{equation}%
where $p$ is the canonical probability using (\ref{Hamiltonian_NF}) and $p_{%
\text{A}}$ is the approximation that we seek. We proceed rewiting the
entropy as 
\begin{equation}
S[p_{\text{A}}|p]=-\sum_{\left\{ s\right\} }p_{\text{A}}\log ~p_{\text{A}%
}+\sum_{\left\{ s\right\} }p_{\text{A}}\log ~p~,
\end{equation}%
or subsituting with the canonical form for $p,$ (\ref{P}) 
\begin{equation}
S[p_{\text{A}}|p]=\frac{1}{k}S_{\text{A}}-\sum_{\left\{ s\right\} }p_{\text{A%
}}\beta H\left\{ s\right\} -\beta F~,
\end{equation}%
where $Z$ can also be written as $Z=e^{-\beta F}$ where $F$ is the free
energy. This can further be reduced to 
\begin{equation}
S[p_{\text{A}}|p]=\frac{1}{k}S_{\text{A}}+\beta \left( F-\left\langle
H\left\{ s\right\} \right\rangle _{\text{A}}\right) ~,
\end{equation}%
where $\left\langle H\left\{ s\right\} \right\rangle _{\text{A}%
}=\sum_{\left\{ s\right\} }p_{\text{A}}H\left\{ s\right\} $ which can also
be seen as the energy $E_{\text{A}}$ of the system. Since by definition, $%
S[p_{\text{A}}|p]\leq 0$ we can write, 
\begin{equation}
F\leq E_{\text{A}}-TS_{\text{A}}~.  \label{Free}
\end{equation}%
When using the ME method, we maximize the entropy to find the best posterior
given the information provided. For this specific case, we have rewritten
the entropy form in terms of an inequality that compares the free energy of
the system, $F$ with the approximate values for the average energy, $E_{%
\text{A}}$ and the entropy $S_{\text{A}}$. The free energy is minimized to
determine the best approximation for $F$, which is called the Bogoliubov
Variational Principle.\cite{Tseng}. However, this is simply a special case
of ME.To be more general, we proceed with using the ME method to find the
best approximation.

When using the ME method, the goal is to search the family of possible
posteriors to find the one that maximizes the entropy given the constraints.
In addition to this, we all need a solution that is trackable. Therefore, we
seek a posterior that has a form, 
\begin{equation}
p_{\text{A}}=\frac{1}{Z_{\text{A}}}e^{-\beta H_{\text{A}}}~,  \label{PA}
\end{equation}%
where $H_{\text{A}}=-\sum_{i}\epsilon _{i\text{A}}s_{i}$ and $\epsilon _{i%
\text{A}}$ is some \emph{effective} energy similar to (\ref{Hamiltonian_effM}%
). Again for illustrative purposes we assume that all atoms poses the same
effective energy so that $\epsilon _{i\text{A}}=\epsilon _{\text{A}}.$ The
difference is that we do not yet \emph{know} the form of $\epsilon _{\text{A}%
}.$ We continue by following a similar route to (\ref{Expected_M}) by
writing the expection value for $s_{i}$ with respect to $p_{\text{A}}$, 
\begin{equation}
\left\langle s_{i}\right\rangle _{\text{A}}=\sum_{\left\{ s_{i}\right\}
}s_{i}~p_{\text{A}}=\tanh \beta \epsilon _{i\text{A}}=\tanh \beta \epsilon _{%
\text{A}}~,  \label{AvgA}
\end{equation}%
Except that here we are \emph{marginalizing} over all atoms except the $%
i^{th}$. Notice that becasue $\epsilon _{\text{A}}$ is constant, the
solution is independent of $i.$ Next we maximize (\ref{Entropy_ME}) or in
keeping with the current case, we minimize the free energy, $F$. Rewiting (%
\ref{Free}) 
\begin{equation}
\beta F_{\min }=-\frac{1}{k}S_{\text{A}}+\beta \left\langle H\left\{
s\right\} \right\rangle _{\text{A}}
\end{equation}%
Subsituting in (\ref{PA}) and rewritting,%
\begin{equation}
F_{\min }=\left( \frac{1}{\beta }\sum_{\left\{ s\right\} }p_{\text{A}}\log
e^{-\beta H_{\text{A}}}-\frac{1}{\beta }\sum_{\left\{ s\right\} }p_{\text{A}%
}\log Z_{\text{A}}\right) +\sum_{\left\{ s\right\} }p_{\text{A}}H\left\{
s\right\} ~,
\end{equation}%
as $Z_{\text{A}}$ is a constant it can be pulled out of the summation and
since $\sum_{\left\{ s\right\} }p_{\text{A}}=1,$ 
\begin{equation}
F_{\min }=\left( -\frac{1}{\beta }\sum_{\left\{ s\right\} }p_{\text{A}}\beta
H_{\text{A}}-\frac{1}{\beta }\log Z_{\text{A}}\right) +\sum_{\left\{
s\right\} }p_{\text{A}}H\left\{ s\right\} 
\end{equation}%
or%
\begin{equation}
F_{\min }=-\left\langle H_{\text{A}}\right\rangle _{\text{A}}-\frac{1}{\beta 
}\log Z_{\text{A}}+\left\langle H\left\{ s\right\} \right\rangle _{\text{A}%
}~.
\end{equation}%
We now substitute our Hamiltonians into the function, 
\begin{equation}
F_{\min }=-\left\langle -\epsilon _{\text{A}}\sum_{i}s_{i}\right\rangle _{%
\text{A}}-\frac{1}{\beta }\log Z_{\text{A}}+\left\langle
-J\sum_{i,j}^{N,n}s_{i}s_{j}-\sum_{i}^{N}\mu Bs_{i}\right\rangle _{\text{A}%
}~,
\end{equation}%
and substituting $\left\langle \sum_{i}s_{i}\right\rangle _{\text{A}%
}=\sum_{i}\left\langle s_{i}\right\rangle _{\text{A}}=N\left\langle
s_{i}\right\rangle _{\text{A}}=N\tanh \beta \epsilon _{\text{A}}$ yields, 
\begin{equation}
F_{\min }=\epsilon _{\text{A}}N\tanh \beta \epsilon _{\text{A}}-\frac{1}{%
\beta }\log Z_{\text{A}}-J\sum_{i,j}^{N,n}\left\langle
s_{i}s_{j}\right\rangle _{\text{A}}-\mu BN\tanh \beta \epsilon _{\text{A}}~.
\end{equation}%
Then substituting $Z_{\text{A}}\approx (2\cosh \beta \epsilon _{\text{A}%
})^{N}$ yields, 
\begin{equation}
F_{\min }=\epsilon _{\text{A}}N\tanh \beta \epsilon _{\text{A}}-\frac{1}{%
\beta }\log (2\cosh \beta \epsilon _{\text{A}})^{N}-J\sum_{i,j}^{N,n}\left%
\langle s_{i}s_{j}\right\rangle _{\text{A}}-\mu BN\tanh \beta \epsilon _{%
\text{A}}
\end{equation}%
and since (\ref{AvgA}) is independent of $i,$ $\left\langle
s_{i}\right\rangle _{\text{A}}=\left\langle s_{j}\right\rangle _{\text{A}}$
and $\left\langle s_{i}s_{j}\right\rangle _{\text{A}}=\left\langle
s_{i}\right\rangle _{\text{A}}\left\langle s_{j}\right\rangle _{\text{A}}$,
we can write, 
\begin{equation}
F_{\min }=\epsilon _{\text{A}}N\tanh \beta \epsilon _{\text{A}}-\frac{1}{%
\beta }N\log (2\cosh \beta \epsilon _{\text{A}})-J\sum_{i,j}^{N,n}\left%
\langle s_{i}\right\rangle _{\text{A}}\left\langle s_{j}\right\rangle _{%
\text{A}}-\mu BN\tanh \beta \epsilon _{\text{A}}~.
\end{equation}%
Substituting (\ref{AvgA}) into the equation yields, 
\begin{equation}
F_{\min }=\epsilon _{\text{A}}N\tanh \beta \epsilon _{\text{A}}-\frac{N}{%
\beta }\log (2\cosh \beta \epsilon _{\text{A}})-J\frac{1}{2}Nn\left( \tanh
\beta \epsilon _{\text{A}}\right) ^{2}-\mu BN\tanh \beta \epsilon _{\text{A~,%
}}
\end{equation}%
where the factor $n$ comes from the number of nearest neighbors, the factor $%
N$ is from the total number of atoms and the 1/2 is due to double counting.

Now we choose the form that minimizes $F_{\min }$ with respect to $\epsilon
_{\text{A}},$ 
\begin{eqnarray}
\frac{\partial F_{\min }}{\partial \epsilon _{\text{A}}} &=&0=N\tanh \beta
\epsilon _{\text{A}}+N\epsilon _{\text{A}}\frac{\beta }{\cosh ^{2}\beta
\epsilon _{\text{A}}}-\frac{N}{\beta }\frac{\beta \sinh \beta \epsilon _{%
\text{A}}}{\cosh \beta \epsilon _{\text{A}}} \\
&&-J\frac{1}{2}Nn2\tanh \beta \epsilon _{\text{A}}\frac{\beta }{\cosh
^{2}\beta \epsilon _{\text{A}}}-\mu BN\frac{\beta }{\cosh ^{2}\beta \epsilon
_{\text{A}}}~,  \notag
\end{eqnarray}%
or rewritten, 
\begin{eqnarray}
0 &=&N\tanh \beta \epsilon _{\text{A}}+N\epsilon _{\text{A}}\frac{\beta }{%
\cosh ^{2}\beta \epsilon _{\text{A}}}-N\tanh \beta \epsilon _{\text{A}} \\
&&-JNn\tanh \beta \epsilon _{\text{A}}\frac{\beta }{\cosh ^{2}\beta \epsilon
_{\text{A}}}-\mu BN\frac{\beta }{\cosh ^{2}\beta \epsilon _{\text{A}}}~. 
\notag
\end{eqnarray}%
After canceling terms we have,  
\begin{equation}
0=\epsilon _{\text{A}}-\mu B-Jn\tanh \beta \epsilon _{\text{A}}
\end{equation}%
or better, 
\begin{equation}
\epsilon _{\text{A}}-\mu B=Jn\tanh \beta \epsilon _{\text{A}}~.  \label{EA}
\end{equation}%
This is our final result for this section. Notice that if $B=0$ then we
recover our solution using the mean field approximation above. If we let $%
\epsilon _{\text{A}}=Jn\bar{s}$ we rewrite the above equation, 
\begin{equation}
Jn\bar{s}=Jn\tanh \beta Jn\bar{s}~,
\end{equation}

and after cancelations, 
\begin{equation}
\bar{s}=\tanh \beta Jn\bar{s}~.
\end{equation}%
which is (\ref{AvgM}).

However, from (\ref{EA})\emph{\ alone} we can solve for the Curie
temperture, the total magnetic moment and the magnetic susceptibility in the
usual way \cite{Caticha}. This is true even though we still do not \emph{know%
} the form for $\epsilon _{\text{A}}.$ Using this approach, we did not need
to assume that all atoms behave like a "central" atom and we did not need to
know the explicit form of the effective energy, only that there was one.
Following the ME method we simply processed all of our information that we
had available. Notice that we also no longer need to \emph{assume} Ising
conditions. We explore such possible scenerios in the next section.

\section{The defective Ising model}

\subsection{Three state example}

In the Ising model, the possible states are spin up and spin down. Let us
examine the case where there are \emph{three} possible states. In this
example, we will let $s=\{+1,0,-1\}$ where an atom in the zero, $0$ state
would contribute no energy, like perhaps a defect of some kind. We proceed
with the method above up until (\ref{AvgA}) where the the new expected value
for a particular atom for this case would be,%
\begin{equation}
\left\langle s_{i}\right\rangle _{\text{A}}=\sum_{\left\{ s_{i}\right\}
}s_{i}~p_{\text{A}}=\frac{2\sinh \beta \epsilon _{\text{A}}}{2\cosh \beta
\epsilon _{\text{A}}+1}~.
\end{equation}%
Continuing the steps above, we write our new $F_{\min },$ 
\begin{equation}
F_{\min }=\epsilon _{\text{A}}N\frac{2\sinh \beta \epsilon _{\text{A}}}{%
2\cosh \beta \epsilon _{\text{A}}+1}-\frac{N}{\beta }\log (2\cosh \beta
\epsilon _{\text{A}}+1)-J\frac{1}{2}Nn\left( \frac{2\sinh \beta \epsilon _{%
\text{A}}}{2\cosh \beta \epsilon _{\text{A}}+1}\right) ^{2}-\mu BN\frac{%
2\sinh \beta \epsilon _{\text{A}}}{2\cosh \beta \epsilon _{\text{A}}+1}~,
\end{equation}%
and minimize this function once again with respect to $\epsilon _{\text{A}}$
which yields,%
\begin{eqnarray}
\frac{\partial F_{\min }}{\partial \epsilon _{\text{A}}} &=&0=N\frac{2\sinh
\beta \epsilon _{\text{A}}}{2\cosh \beta \epsilon _{\text{A}}+1}+N\epsilon _{%
\text{A}}\left( \frac{\beta 2\cosh \beta \epsilon _{\text{A}}}{2\cosh \beta
\epsilon _{\text{A}}+1}-\frac{\beta 4\sinh ^{2}\beta \epsilon _{\text{A}}}{%
\left( 2\cosh \beta \epsilon _{\text{A}}+1\right) ^{2}}\right) -\frac{N}{%
\beta }\frac{2\beta \sinh \beta \epsilon _{\text{A}}}{2\cosh \beta \epsilon
_{\text{A}}+1} \\
&&-J\frac{1}{2}Nn2\left( \frac{2\sinh \beta \epsilon _{\text{A}}}{2\cosh
\beta \epsilon _{\text{A}}+1}\right) \left( \frac{\beta 2\cosh \beta
\epsilon _{\text{A}}}{2\cosh \beta \epsilon _{\text{A}}+1}-\frac{\beta
4\sinh ^{2}\beta \epsilon _{\text{A}}}{\left( 2\cosh \beta \epsilon _{\text{A%
}}+1\right) ^{2}}\right) -\mu BN\left( \frac{\beta 2\cosh \beta \epsilon _{%
\text{A}}}{2\cosh \beta \epsilon _{\text{A}}+1}-\frac{\beta 4\sinh ^{2}\beta
\epsilon _{\text{A}}}{\left( 2\cosh \beta \epsilon _{\text{A}}+1\right) ^{2}}%
\right) ~.
\end{eqnarray}%
After canceling terms we arrive at, 
\begin{equation}
\epsilon _{\text{A}}-\mu B=Jn\left( \frac{2\sinh \beta \epsilon _{\text{A}}}{%
2\cosh \beta \epsilon _{\text{A}}+1}\right) .
\end{equation}%
Which is our final result. This equation can then be used as above to solve
for the Curie temperture, the total magnetic moment and the magnetic
susceptibility for this material.

To make this result a little more genreral we can also write, 
\begin{equation}
\epsilon _{\text{A}}-\mu B=Jn_{\text{G}}\left( -\frac{\partial \ln Z_{\text{A%
}}}{\partial \epsilon _{\text{A}}}\right) =Jn_{\text{G}}\beta \left( \frac{%
\partial F_{\text{A}}}{\partial \epsilon _{\text{A}}}\right) =Jn_{\text{G}%
}\left( \left\langle s_{i}\right\rangle _{\text{A}}\right) ~,  \label{EA_G}
\end{equation}%
where $Z_{\text{A}}$ is once again the \emph{approximate} partition
function, $n_{\text{G}}$ are the nearest neighbors and $F_{\text{A}}$ is the 
\emph{approximate} free energy.

\subsection{Two energy example}

In the previous example, we examined a three state atom. In this example, we
examine a material that has an \emph{known} defect. Perhaps the material was
scanned in some way and a defect was found. In this case, we need to use%
\emph{\ two} effective energy terms. One for the atoms that are surounded by
non-defective atoms, $\epsilon _{\text{ND}}$ and one for the ones affected
by the defect, $\epsilon _{\text{D}}.$ For illustrative purposes, we will
examine the case where we have one defective atom. This means that there are 
$N-n-1$ atoms that are surrounded by non-defective atoms, $n$ atoms that
have one defective atom next to it and $1$ atom which is the defective atom.
For our purposes, let's think of the defect as an empty slot. Therefore, the
actual Hamiltonian would be,%
\begin{equation*}
H\left\{ s\right\}
=H_{int}+H_{ext}=-J\sum_{i,j}^{N-n-1,n}s_{i}s_{j}-J%
\sum_{k,l}^{n,n-1}s_{k}s_{l}-Js_{0}-\sum_{i}^{N-n-1}\mu
Bs_{i}-\sum_{k}^{n}\mu Bs_{k}-\mu Bs_{0}~,
\end{equation*}%
where $i$ are the non-defective atoms, $j$ are neighbors ($n$) for these
atoms, $k$ are the atoms affected by the defect, $l$ are the neighbors ($n-1)
$ of the affected atoms and $s_{0}$ is the defective atom. Since we are
looking at this as an empty slot, we let $s_{0}=0$. Now we write our
estimated Hamiltonian for this case, 
\begin{equation*}
H_{\text{A}}=-\epsilon _{\text{ND}}\sum_{i}s_{i}-\epsilon _{\text{D}%
}\sum_{k}s_{k}~.
\end{equation*}%
Following the same procedures above, we attain \emph{two} expected values,
one for each effective energy, 
\begin{equation}
\left\langle s_{i}\right\rangle _{\text{A}}=\sum_{\left\{ s_{i}\right\}
}s_{i}~p_{\text{A}}=\tanh \beta \epsilon _{\text{ND}}~  \label{S_ND}
\end{equation}%
and%
\begin{equation}
\left\langle s_{k}\right\rangle _{\text{A}}=\sum_{\left\{ s_{i}\right\}
}s_{k}~p_{\text{A}}=\tanh \beta \epsilon _{\text{D}}~.  \label{S_D}
\end{equation}%
Once again we write down the function we wish to minimize, 
\begin{equation}
F_{\min }=-\left\langle H_{\text{A}}\right\rangle _{\text{A}}-\frac{1}{\beta 
}\log Z_{\text{A}}+\left\langle H\left\{ s\right\} \right\rangle _{\text{A}%
}~,
\end{equation}%
and substituting in our Hamiltonians yields, 
\begin{equation}
F_{\min }=-\left\langle -\epsilon _{\text{ND}}\sum_{i}^{N-n-1}s_{i}-\epsilon
_{\text{D}}\sum_{k}^{n}s_{k}\right\rangle _{\text{A}}-\frac{1}{\beta }\log
Z_{\text{A}}+\left\langle
-J\sum_{i,j}^{N-n-1,n}s_{i}s_{j}-J\sum_{k,l}^{n,n-1}s_{k}s_{l}-%
\sum_{i}^{N-n-1}\mu Bs_{i}-\sum_{k}^{n}\mu Bs_{k}\right\rangle _{\text{A}}~.
\end{equation}%
Substituting $\left\langle \sum_{i}s_{i}\right\rangle _{\text{A}%
}=\sum_{i}\left\langle s_{i}\right\rangle _{\text{A}}=\left( N-n-1\right)
\left\langle s_{i}\right\rangle _{\text{A}}=\left( N-n-1\right) \tanh \beta
\epsilon _{\text{ND}}$ and $\left\langle \sum_{k}s_{k}\right\rangle _{\text{A%
}}=\sum_{k}\left\langle s_{k}\right\rangle _{\text{A}}=n\left\langle
s_{k}\right\rangle _{\text{A}}=n\tanh \beta \epsilon _{\text{D}}$ yields, 
\begin{eqnarray}
F_{\min } &=&\epsilon _{\text{ND}}\left( N-n-1\right) \tanh \beta \epsilon _{%
\text{ND}}+\epsilon _{\text{D}}n\tanh \beta \epsilon _{\text{D}}-\frac{1}{%
\beta }\log \left( Z_{\text{ND}}Z_{\text{D}}\right)  \\
&&-J\sum_{i,j}^{N-n-1,n}\left\langle s_{i}s_{j}\right\rangle _{\text{ND}%
}-J\sum_{k,l}^{n,n-1}\left\langle s_{k}s_{l}\right\rangle _{\text{D}}-\mu B%
\left[ \epsilon _{\text{ND}}\left( N-n-1\right) \tanh \beta \epsilon _{\text{%
ND}}+\epsilon _{\text{D}}n\tanh \beta \epsilon _{\text{D}}\right] ~,
\end{eqnarray}%
where $Z_{\text{A}}=Z_{\text{ND}}Z_{\text{D}}.$ Substituting $Z_{\text{ND}%
}\approx (2\cosh \beta \epsilon _{\text{ND}})^{\left( N-n-1\right) }$ and $%
Z_{\text{D}}\approx (2\cosh \beta \epsilon _{\text{D}})^{\left( n\right) }$
yields, 
\begin{eqnarray}
F_{\min } &=&\epsilon _{\text{ND}}\left( N-n-1\right) \tanh \beta \epsilon _{%
\text{ND}}+\epsilon _{\text{D}}n\tanh \beta \epsilon _{\text{D}}-\frac{1}{%
\beta }\log (2\cosh \beta \epsilon _{\text{ND}})^{N-n-1}-\frac{1}{\beta }%
\log (2\cosh \beta \epsilon _{\text{D}})^{n-1} \\
&&-J\sum_{i,j}^{N-n-1,n}\left\langle s_{i}s_{j}\right\rangle _{\text{ND}%
}-J\sum_{k,l}^{n,n-1}\left\langle s_{k}s_{l}\right\rangle _{_{\text{D}}}-\mu
B\left[ \epsilon _{\text{ND}}\left( N-n-1\right) \tanh \beta \epsilon _{%
\text{ND}}+\epsilon _{\text{D}}n\tanh \beta \epsilon _{\text{D}}\right] ~.
\end{eqnarray}%
Since (\ref{S_ND}) and (\ref{S_D}) are independent of $i$ and $k$
respectively,  $\left\langle s_{i}\right\rangle _{\text{A}}=\left\langle
s_{j}\right\rangle _{\text{A}}$ and $\left\langle s_{i}s_{j}\right\rangle _{%
\text{A}}=\left\langle s_{i}\right\rangle _{\text{A}}\left\langle
s_{j}\right\rangle _{\text{A}}$, we can write, 
\begin{eqnarray}
F_{\min } &=&\epsilon _{\text{ND}}\left( N-n-1\right) \tanh \beta \epsilon _{%
\text{ND}}+\epsilon _{\text{D}}\left( n\right) \tanh \beta \epsilon _{\text{D%
}}-\frac{1}{\beta }\log (2\cosh \beta \epsilon _{\text{ND}})^{N-n-1}-\frac{1%
}{\beta }\log (2\cosh \beta \epsilon _{\text{D}})^{n-1} \\
&&-J\sum_{i,j}^{N-n-1,n}\left\langle s_{i}\right\rangle _{\text{ND}%
}\left\langle s_{j}\right\rangle _{\text{ND}}-J\sum_{k,l}^{n,n-1}\left%
\langle s_{k}\right\rangle _{\text{D}}\left\langle s_{l}\right\rangle _{%
\text{D}}-\mu B\left[ \epsilon _{\text{ND}}\left( N-n-1\right) \tanh \beta
\epsilon _{\text{ND}}+\epsilon _{\text{D}}\left( n\right) \tanh \beta
\epsilon _{\text{D}}\right] ~.
\end{eqnarray}%
Substituting (\ref{S_ND}) and (\ref{S_D}) into the equation yields, 
\begin{eqnarray}
F_{\min } &=&\epsilon _{\text{ND}}\left( N-n-1\right) \tanh \beta \epsilon _{%
\text{ND}}+\epsilon _{\text{D}}\left( n\right) \tanh \beta \epsilon _{\text{D%
}}-\frac{1}{\beta }\log (2\cosh \beta \epsilon _{\text{ND}})^{N-n-1}-\frac{1%
}{\beta }\log (2\cosh \beta \epsilon _{\text{D}})^{n-1} \\
&&-\frac{1}{2}J\left( N-n-1\right) \left( n\right) \tanh ^{2}\beta \epsilon
_{\text{ND}}-\frac{1}{2}J\left( n\right) \left( n-1\right) \tanh ^{2}\beta
\epsilon _{\text{D}}-\mu B\left[ \epsilon _{\text{ND}}\left( N-n-1\right)
\tanh \beta \epsilon _{\text{ND}}+\epsilon _{\text{D}}\left( n\right) \tanh
\beta \epsilon _{\text{D}}\right] 
\end{eqnarray}%
and after collecting $\epsilon _{\text{ND}}$ and $\epsilon _{\text{D}}$ like
terms we have, 
\begin{eqnarray*}
F_{\min } &=&\epsilon _{\text{ND}}\left( N-n-1\right) \tanh \beta \epsilon _{%
\text{ND}}-\frac{1}{\beta }\log (2\cosh \beta \epsilon _{\text{ND}})^{N-n-1}-%
\frac{1}{2}J\left( N-n-1\right) \left( n\right) \tanh ^{2}\beta \epsilon _{%
\text{ND}}\mu B\epsilon _{\text{ND}}\left( N-n-1\right) \tanh \beta \epsilon
_{\text{ND}} \\
&&+\epsilon _{\text{D}}\left( n\right) \tanh \beta \epsilon _{\text{D}}-%
\frac{1}{\beta }\log (2\cosh \beta \epsilon _{\text{D}})^{n-1}-\frac{1}{2}%
J\left( n\right) \left( n-1\right) \tanh ^{2}\beta \epsilon _{\text{D}}-+\mu
B\epsilon _{\text{D}}\left( n\right) \tanh \beta \epsilon _{\text{D}}~,
\end{eqnarray*}%
where the factor $n$ comes from the number of nearest neighbors, the factor $%
N$ is from the total number of atoms and the 1/2 is due to double counting.

Now we choose the form that minimizes $F_{\min }$ with respect to $\epsilon
_{\text{ND}}$ and $\epsilon _{\text{D}},$ 
\begin{eqnarray}
\frac{\partial F_{\min }}{\partial \epsilon _{\text{ND}}} &=&0=\left(
N-n-1\right) \tanh \beta \epsilon _{\text{ND}}+\left( N-n-1\right) \epsilon
_{\text{ND}}\frac{\beta }{\cosh ^{2}\beta \epsilon _{\text{ND}}}-\frac{N-n-1%
}{\beta }\frac{\beta \sinh \beta \epsilon _{\text{ND}}}{\cosh \beta \epsilon
_{\text{ND}}} \\
&&-J\frac{1}{2}\left( N-n-1\right) \left( n\right) 2\tanh \beta \epsilon _{%
\text{ND}}\frac{\beta }{\cosh ^{2}\beta \epsilon _{\text{ND}}}-\mu B\left(
N-n-1\right) \frac{\beta }{\cosh ^{2}\beta \epsilon _{\text{ND}}}  \notag
\end{eqnarray}%
\begin{eqnarray}
0 &=&\left( N-n-1\right) \tanh \beta \epsilon _{\text{ND}}+\left(
N-n-1\right) \epsilon _{\text{ND}}\frac{\beta }{\cosh ^{2}\beta \epsilon _{%
\text{ND}}}-\frac{N-n-1}{\beta }\frac{\beta \sinh \beta \epsilon _{\text{ND}}%
}{\cosh \beta \epsilon _{\text{ND}}} \\
&&-J\left( N-n-1\right) \left( n\right) \tanh \beta \epsilon _{\text{ND}}%
\frac{\beta }{\cosh ^{2}\beta \epsilon _{\text{ND}}}-\mu B\left(
N-n-1\right) \frac{\beta }{\cosh ^{2}\beta \epsilon _{\text{ND}}}  \notag
\end{eqnarray}%
\begin{equation*}
0=\epsilon _{\text{ND}}-J\left( n\right) \tanh \beta \epsilon _{\text{ND}%
}-\mu B
\end{equation*}%
\begin{equation}
\epsilon _{\text{ND}}-\mu B=Jn\tanh \beta \epsilon _{\text{ND}}
\end{equation}%
\begin{equation*}
\epsilon _{\text{D}}-\mu B=J\left( n-1\right) \tanh \beta \epsilon _{\text{D}%
}
\end{equation*}%
Which is our final result. These equations can then be used as above to
solve for the Curie temperture, the total magnetic moment and the magnetic
susceptibility for this material as usual. The difference here is that we
will have two of each. For example, a Curie temperture that applies to most
of the atoms, and one that is local to the defect. Therefore, the total
magnetic moment will not simple be a result of just one tempreture but a sum
of each magnetic moment determined by the equations. Notice that (\ref{EA_G}%
) still holds in general. In the two energy case, for one energy the number
of nearest neighbors, $n_{\text{G}}=n$ and for the second energy, $n_{\text{G%
}}=n-1.$

\section{Conclusions}

We described the traditional Ising model of a ferromagnetic material and the
mean field approximation for a multidimentional material. It has been shown
that not only can the Me method be used for updateing, but for determining
approximations as well.

\noindent \textbf{Acknowledgements:} I would like to acknowledge valuable
discussions with Ariel Caticha regarding this material.

\end{document}